# Stellar control on atmospheric carbon chemistry, CO runaway, and organic synthesis on lifeless Earth-like planets

Yoshiaki Endo[1,2], Yasuto Watanabe[3], Kazumi Ozaki[2,4]

**Abstract**

The abundances of atmospheric carbon species—carbon dioxide ($CO_2$), carbon monoxide (CO), and methane ($CH_4$)—exert fundamental controls on the climate, redox state, and prebiotic environment of terrestrial planets. As exoplanet atmospheric characterization advances, it is essential to understand how these species are regulated on habitable terrestrial planets across a wide range of stellar and planetary conditions. Here, we develop an integrated numerical model that couples atmospheric chemistry, climate, and the long-term carbon cycle to investigate the atmospheric compositions of lifeless, Earth-like planets orbiting Sun-like (F-, G-, and K-type) stars. Our simulations demonstrate that $CO_2$, CO, and $CH_4$ generally increase with orbital distance, and that planets near the outer edge of the habitable zone may undergo CO runaway—a photochemical instability driven by severe depletion of OH radicals. The threshold for CO runaway depends strongly on stellar spectral type and is most easily triggered around cooler, lower-mass stars. In contrast, the atmospheric production of formaldehyde ($H_2CO$)—a key precursor for prebiotic organic chemistry—peaks around planets orbiting more massive, UV-luminous stars and is maximized at orbital distances just interior to the CO-runaway threshold. These results establish a quantitative framework linking observable system properties—stellar type and orbital distance—and the atmospheric carbon chemistry of lifeless Earth-like planets, providing new context for interpreting future spectroscopic observations and for evaluating the potential of such planets to sustain prebiotic chemistry.

[1] Corresponding author yendo@eps.sci.isct.ac.jp
[2] Department of Earth and Planetary Sciences, Institute of Science Tokyo, Ookayama 2-12-1, Meguro-ku, Tokyo 152-8551, Japan
[3] Earth System Division, National Institute for Environmental Studies, 16-2 Onogawa, Tsukuba, Ibaraki 305-8506, Japan
[4] Earth-Life Science Institute (ELSI), Institute of Science Tokyo, Ookayama 2-12-1, Meguro-ku, Tokyo 152-8550, Japan

1. **Introduction**

Since the first detection of an exoplanet in 1995 (Mayor & Queloz 1995), more than 6,000 exoplanets have been confirmed (NASA Exoplanet Archive, accessed 2025 September 29), and the field of exoplanet astronomy is gradually transitioning from the discovery of new planets to the characterization of their environments. Recent advances—most notably with the James Webb Space Telescope (JWST)—are beginning to reveal key atmospheric constituents including $H_2O$, $CO_2$, and $CH_4$ for a growing number of exoplanets (e.g., JWST Transiting Exoplanet Community Early Release Science Team 2023; Bell et al. 2023; Tsai et al. 2023; Hu et al. 2024). Proposed missions such as NASA's Habitable Worlds Observatory (HWO) aim to characterize the atmospheres of Earth-sized habitable planets orbiting Sun-like stars and to search for atmospheric biosignatures in the coming decades. Interpreting such observations requires a mechanistic understanding of the physical and chemical processes that regulate the climates and atmospheric compositions of terrestrial planets.

The classical concept of the habitable zone (HZ) defines the circumstellar region where a terrestrial planet with a $CO_2$–$H_2O$–$N_2$ atmosphere can maintain surface liquid water (e.g., Huang 1959; Hart 1979; Kasting et al. 1993; Kopparapu et al. 2013). Although dozens of exoplanets have already been identified within the HZ (Anglada-Escudé et al. 2016; Gillon et al. 2017; Hill et al. 2023), their actual ability to sustain life depends on a variety of environmental factors, including atmospheric composition, volcanic activity, and the global carbon cycle (Cockell et al. 2016; Ehlmann et al. 2016). Among these factors, the atmospheric carbon-bearing species—specifically the relative abundances of $CO_2$, $CO$, and $CH_4$—play particularly important roles in regulating climate, ocean chemistry, and redox conditions relevant to prebiotic environments. $CO_2$ and $CH_4$ are potent greenhouse gases (e.g., Owen et al. 1979; Kiehl & Dickinson 1987), while CO serves as a key intermediate in atmospheric chemistry and a precursor for prebiotic organic synthesis (e.g., Abelson 1966; Pinto et al. 1980; Miyakawa et al. 2002). Despite their importance, most previous theoretical studies have focused on $CO_2$, because of its central role in stabilizing climate through the carbonate-silicate geochemical cycle (e.g., Walker et al. 1981; Tajika & Matsui, 1992; Kasting et al. 1993; Sleep & Zahnle 2001; Kadoya & Tajika, 2014, 2019; Menou 2015; Bean et al. 2017; Lehmer et al. 2020; Affholder et al. 2025). Much less is understood about what determines the relative abundances of $CO_2$, $CO$, and $CH_4$ in the atmospheres of Earth-like planets within the HZ, or what trends these carbon species may exhibit across orbital distances.

CO is of particular interest because of its dual relevance to both prebiotic chemistry and biosignature interpretation. Laboratory experiments and theoretical studies demonstrate that CO serves as a precursor to essential prebiotic feedstock molecules, such as formaldehyde ($H_2CO$), and may also have mediated primitive metabolic pathways on the early Earth (e.g., Pinto et al., 1980; Zang et al., 2022). Moreover, atmospheric $CO/CH_4$ ratio is increasingly recognized as a key diagnostic in assessing biosignature false positives (Schwieterman et al. 2019; Wogan and Catling 2020; Thompson et al. 2022; Ozaki et al. in press). The relative abundances of $CO_2$, CO, and $CH_4$ therefore encode information not only about planetary climate but also about the availability of chemical substrates and energy sources that could support the prebiotic chemistry, as well as about atmospheric chemical states that may help distinguish inhabited from uninhabited worlds.

Photochemical modeling has revealed that CO can even enter a runaway regime in anoxic atmospheres, where OH radicals—primarily produced by $H_2O$ photolysis—become severely depleted (Kasting et al. 1983; Zahnle 1986; Ranjan et al. 2022; Watanabe & Ozaki 2024). This CO runaway regime is favored under (1) high atmospheric $CO_2$ levels, (2) strong outgassing fluxes of reducing gases, (3) cool climates with low $H_2O$ vapor, and (4) low UV flux from low-mass stars (e.g., Watanabe & Ozaki 2024). Conditions (1) and (3) are broadly consistent with planets located near the outer edge of the HZ, yet it remains unclear whether the CO runaway threshold exists in the HZs around Sun-like stars, and how $CO_2$, CO, and $CH_4$ jointly vary across lifeless Earth-like planets within the HZ.

Furthermore, the importance of the atmospheric carbon redox spectrum ($CO_2$–CO–$CH_4$) for prebiotic chemistry remains unclear. Formaldehyde ($H_2CO$), in particular, is a key precursor for sugar synthesis via the formose reaction (e.g., Cleaves 2008), and its atmospheric production and delivery to the surface environments depend strongly on the atmospheric chemistry (e.g., Pinto et al. 1980; Zang et al. 2022; Watanabe & Ozaki 2024). Identifying where in orbital distance space $H_2CO$ production is maximized—and how this position shifts with stellar type—offers a pathway for connecting planetary photochemistry to potential prebiotic environments.

In this study, we investigate how atmospheric $CO_2$, CO, and $CH_4$ abundances vary across lifeless, Earth-like terrestrial planets orbiting F-, G-, and K-type stars, focusing on the interplay of photochemistry, climate, and the long-term carbon cycle. Here we define Earth-like planets as prebiotic Earth analogues that possess global oceans, active volcanism, and carbonate-silicate geochemical cycle, but lack biological activity. Using an integrated atmosphere-climate-carbon cycle model, we quantify the steady-state abundances of $CO_2$, CO, and $CH_4$ in the atmosphere and identify the planetary and stellar conditions that lead to CO runaway. We further examine the atmospheric production and surface deposition of $H_2CO$ to assess potential links between atmospheric compositions and prebiotic organic synthesis.

## 2. Methods
### 2.1. Overall model design

Our modeling framework integrates three components: a 1-D atmospheric photochemical model, a coupled climate-carbon cycle model, and a volcanic outgassing model (Figure 1). These submodels interact to determine the steady-state atmospheric composition of lifeless Earth-like planets.

The core of the model is the vertically resolved 1-D photochemical model *Atmos* (Arney et al. 2016; Section 2.2), which computes the atmospheric chemical structure. Its lower and upper boundary conditions include the UV radiation spectrum, surface fluxes of volcanic gases, the atmospheric $pCO_2$, and the surface temperature ($T_s$). The UV flux is set by the host-star spectral type (F-, G-, and K-type) and the orbital semi-major axis, assuming circular orbits.

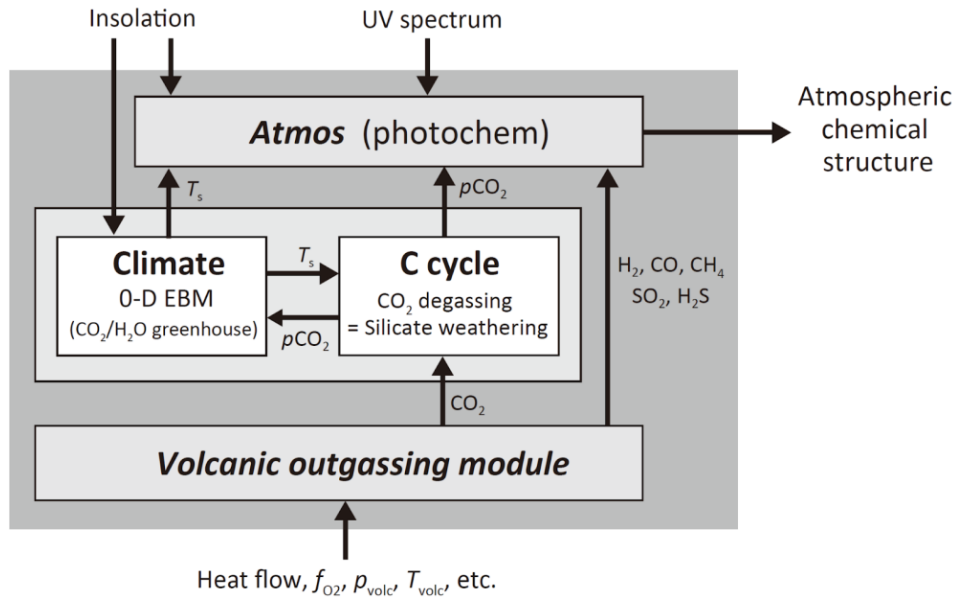

Figure 1. Schematic of model structure illustrating the relationships between submodels and key external forcings examined in this study.

Surface fluxes of reducing gases are calculated using the thermodynamic equilibrium model of Wogan & Catling (2020) (Section 2.3), which calculates the equilibrium composition of volcanic gases at specified pressure ($p_{volc}$), temperature ($T_{volc}$), oxygen fugacity ($f_{O2}$), and elemental H/C/S ratios. The resulting molar fractions of $H_2O$, $H_2$, $CO_2$, $CO$, $CH_4$, $SO_2$, and $H_2S$ are then converted to global outgassing fluxes scaling with a prescribed seafloor spreading rate, $r_{spr}$ (Section 2.3). The volcanic $CO_2$ flux is passed to the carbon-cycle model (Section 2.4.2), whereas fluxes of $H_2$, $CO$, $CH_4$, $SO_2$, and $H_2S$ are used as lower boundary conditions for *Atmos* (Section 2.2). The coupled climate-carbon cycle model (Section 2.4) determines atmospheric $pCO_2$ and $T_s$. The climate model is a 0-D energy balance model (EBM) that evaluates the energy balance based on incoming stellar radiation and outgoing longwave radiation. The carbon cycle model computes the mass balance of $CO_2$ in the ocean-atmosphere system through volcanic outgassing and silicate weathering on land and the seafloor.

Given these boundary conditions, the photochemical model is integrated to steady state. To improve numerical convergence, each experiment is repeated three times, using the output of the previous run as the initial condition (Watanabe & Ozaki 2024; Akahori et al. 2024). At steady state, the net oxidizing and reducing power fluxes in the atmosphere become balanced. In our model, atmospheric climate is controlled primarily by $CO_2$ (and $H_2O$), which dominate the greenhouse effect under the conditions explored. Because $CH_4$ remains a minor constituent and contributes negligibly to radiative forcing in our simulations, we neglect feedbacks from photochemically produced trace species to climate.

In this study, we perform a suite of sensitivity experiments by varying four key parameters: (1) the stellar effective flux, $S_{eff}$, (2) the seafloor spreading rate, $r_{spr}$, (3) the stellar spectral type (F-, G-, and K-type), and (4) the oxygen fugacity of volcanic gases, $f_{O2}$ (Section 2.5), where $S_{eff}$ is defined as the stellar insolation normalized by the present solar constant ($S^{PD} = 1361$ W m$^{-2}$; Kopp & Lean 2011).

## 2.2. Atmospheric photochemical model, Atmos

We employ the version of Atmos used by Watanabe & Ozaki (2024), with modifications to boundary conditions for this study. The model resolves the atmosphere from the surface to 100 km using 200 vertical layers, and adopts the eddy diffusion coefficient used in Pavlov et al. (2001).

The surface temperature, $T_s$, and atmospheric $p$CO$_2$ are prescribed by the coupled climate–carbon cycle model (Section 2.4). The vertical temperature profile is then calculated following Watanabe & Ozaki (2024) with a tropospheric lapse rate of 8 K km$^{-1}$. Above the altitude at which the air temperature reaches 180 K, we assume an isothermal air temperature of 180 K. Volcanic fluxes of H$_2$, CO, CH$_4$, SO$_2$, and H$_2$S are supplied by the outgassing model described in Section 2.3. The incident stellar radiation is also imposed as an external forcing. The UV spectrum at the top of the atmosphere is determined by the stellar spectral type (Section 2.5).

### 2.3. Volcanic outgassing model

The global volcanic outgassing flux of each gas species is defined as the sum of subaerial and submarine contributions:

$$F_X = F_X^{\text{subaerial}} + F_X^{\text{submarine}}, \quad (1)$$

where $X$ denotes an individual molecular species (H$_2$O, H$_2$, CO$_2$, CO, CH$_4$, SO$_2$, and H$_2$S). In this study, volcanic gas compositions are calculated using the thermodynamic equilibrium model of Wogan & Catling (2020). For each volcanic setting, the equilibrium composition of volcanic gases is computed as a function of the volcanic pressure ($p_{\text{volc}}$), temperature ($T_{\text{volc}}$), oxygen fugacity ($f_{\text{O2}}$), total hydrogen ($\Sigma$H = 2×H$_2$ + 2×H$_2$O + 4×CH$_4$ + 2×H$_2$S), total carbon ($\Sigma$C = CO$_2$ + CO + CH$_4$), and total sulfur ($\Sigma$S = SO$_2$ + H$_2$S).

Among these parameters, $p_{\text{volc}}$ exerts a profound control on the speciation of reducing materials (e.g., Gaillard et al. 2011; Wogan et al. 2020). Following Wogan & Catling (2020), we adopt $p_{\text{volc}}$ = 5 bar for subaerial volcanoes and 400 bar for submarine volcanoes, consistent with present-day Earth conditions (Holland 1984). $T_{\text{volc}}$ is fixed at 1473 K for both settings (Holland 1984; Wogan & Catling 2020). Oxygen fugacity, $f_{\text{O2}}$, is assumed to follow the fayalite–magnetite–quartz (FMQ) buffer as the standard condition, and we additionally explore a more reducing condition of FMQ−1 (see Appendix B). The relative elemental ratios of hydrogen, carbon, and sulfur (e.g., $\Sigma$H/$\Sigma$C/$\Sigma$S) are assumed to be identical to those of modern Earth subaerial or submarine volcanism, respectively. For simplicity, the effect of atmospheric pressure—determined by the climate–carbon cycle model (Section 2.4)—on equilibrium speciation is neglected.

To assign absolute fluxes, we assume that the total elemental outgassing fluxes of carbon, hydrogen, and sulfur from both subaerial and submarine volcanism are proportional to the relative seafloor spreading rate, $r_{\text{spr}}$, normalized to the present-day Earth value. This assumption follows long-term carbon cycle modeling approaches commonly applied to the Phanerozoic (e.g., Berner et al. 1983; Tajika 1998). Under this assumption, the subaerial and submarine elemental flux of element $X$ ($X = \Sigma H$, $\Sigma C$, or $\Sigma S$) is given by

$$F_X^{\text{subaerial}} = r_{\text{spr}}\, F_X^{\text{subaerial,PD}}, \quad (2)$$

$$F_X^{\text{submarine}} = r_{\text{spr}}\, F_X^{\text{submarine,PD}}, \quad (3)$$

where $F_X^{\text{subaerial,PD}}$ and $F_X^{\text{submarine,PD}}$ are present-day Earth estimates taken from Catling & Kasting (2017). Given the equilibrium-derived relative abundances of individual gas species and the elemental fluxes, the absolute molecular fluxes of volcanic gases are determined by requiring that (i) the relative abundances of reduced and oxidized species follow thermodynamic equilibrium and (ii) the stoichiometrically weighted sums of carbon-, hydrogen-, and sulfur-bearing species equal the prescribed elemental fluxes (Wogan & Catling 2020, Equations A14–A20). This procedure uniquely determines the molecular fluxes while conserving the total inventories of carbon, hydrogen, and sulfur.

### 2.4. The coupled climate-carbon cycle model

Given the stellar irradiance and $CO_2$ outgassing flux, the steady-state values of $T_s$ and $pCO_2$ are estimated by the coupled climate–carbon cycle model. The energy balance model was run until it reaches equilibrium, assuming the carbon mass balance in the ocean-atmosphere system. Under present Earth conditions, the atmospheric $pCO_2$ is $2.80 \times 10^{-4}$ bar, $T_s$ is 288 K, and the climate sensitivity is 4.7 K.

### 2.4.1. Climate model

For the climate model, we use a 0-D EBM (e.g., Pierrehumbert et al. 2011; Menou 2015; Ozaki & Reinhard 2021). The time evolution of $T_s$ can be written as follows:

$$C\frac{dT_s}{dt} = J_{ISR} - J_{OLR}, \quad (4)$$

where $C$ denotes the heat capacity of the atmosphere and the ocean surface layer ($10^{10}$ J m$^{-2}$ K$^{-1}$), $J_{ISR}$ and $J_{OLR}$ represent the incoming solar radiation and the outgoing longwave radiation, respectively. $J_{ISR}$ is given by:

$$J_{ISR} = \frac{S}{4}(1 - a), \quad (5)$$

where $S$ represents the solar constant (= $S_{eff} \times S^{PD}$). The planetary albedo ($a$) is determined using the parametric expressions provided in Appendix A.1 of Haqq-Misra et al. (2016). These expressions depend on the stellar spectral type, $T_s$, atmospheric $pCO_2$, solar zenith angle ($\theta$; $\cos \theta$ = 0.45), and surface albedo ($a_s$). $a_s$ is calculated as the area-weighted mean of the albedos of ocean ($a_{sea}$), land ($a_{land}$), ice ($a_{ice}$), and clouds ($a_{cloud}$) (e.g., Kadoya & Tajika 2019; Rosing et al. 2010) (see Appendix A for details). $J_{OLR}$ is calculated using the parametric expressions in Appendix E of Williams & Kasting (1997), which are functions of $T_s$ and $pCO_2$ and account for the greenhouse effects of $CO_2$ and water vapor. The parametric expressions are valid for atmospheric $pCO_2$ between $10^{-5}$ and 10 bar. Other greenhouse gases (such as $CH_4$) are not considered in the model.

### 2.4.2. Carbon cycle model

The mass balance of $CO_2$ in the ocean-atmosphere system can be written as follows:

$$F_{CO2} = F_{ws} + F_{sws}, \quad (6)$$

where $F_{CO2}$ represents $CO_2$ outgassing, while $F_{ws}$ and $F_{sws}$ denote silicate weathering on land and the seafloor (e.g., Berner 2004; Krissansen-Totton & Catling 2017), respectively. $F_{CO2}$ is obtained from the outgassing model (Section 2.3). $F_{ws}$ and $F_{sws}$ are functions of atmospheric $pCO_2$, $T_s$, $f_{cont}$ (the continental area fraction), and pore-water chemistry (Table 1), and Equation (6) is numerically solved for $pCO_2$. The parameter values are summarized in Table 2.

Table 1. Equations for continental and seafloor weathering fluxes.

| Eq. No | Description | Equation* | Unit | Ref. |
|---|---|---|---|---|
| 7 | Continental silicate weathering ($F_{ws}$) | $F_{ws} = F_{ws}^{PD} \left(\dfrac{f_{cont}}{f_{cont}^{PD}}\right) f_e \left(\dfrac{pCO_2}{pCO_2^{PD}}\right)^{\alpha} \exp\left[-\dfrac{E_{cont}}{R}\left(\dfrac{1}{T_s} - \dfrac{1}{T_s^{PD}}\right)\right]$ | Tmol C yr$^{-1}$ | (1) |
| 8 | Seafloor weathering rate ($F_{sws}$) | $F_{sws} = F_{sws}^{PD} r_{spr} \left(\dfrac{[H^+]_{pore}}{[H^+]_{pore}^{PD}}\right)^{\gamma} \exp\left[-\dfrac{E_{diss}}{R}\left(\dfrac{1}{T_{pore}} - \dfrac{1}{T_{pore}^{PD}}\right)\right]$ | Tmol C yr$^{-1}$ | (2) |
| 9 | Hydrogen ion concentration in the pore space of the oceanic crust ($[H^+]_{pore}$) | $[H^+]_{pore} = (10^{-8.44} \times pCO_2)^{1/1.34}$ | mol L$^{-1}$ | (2) |
| 10 | Deep-sea temperature ($T_d$) | $T_d = 1.1\, T_s - 39.463$ | K | (2) |
| 11 | Pore-space temperature ($T_{pore}$) | $T_{pore} = 1.1\, T_d + 9.0$ | K | (2) |

References. (1) Watanabe & Tajika (2021); (2) Krissansen-Totton et al. (2018a).

*"PD" denotes present-day Earth reference values.

Table 2. Model parameters used in this study.

| Eq. No. | Parameter | Description | Value | Unit | Ref. |
|---|---|---|---|---|---|
| 7 | $F_{CO2}^{PD}$ | Reference value of global $CO_2$ outgassing | 8.5 | Tmol yr$^{-1}$ | (1) |
| 7 | $f_{cont}$ | Continental area fraction | 0.075 | | (2) |
| 7 | $f_{cont}^{PD}$ | Reference continental area fraction (present-day Earth) | 0.3 | | |
| 7 | $f_e$ | Efficiency of soil biological activity (relative to present) | 0.15 | | (3) |
| 7 | $\alpha$ | $pCO_2$ dependence exponent for continental silicate weathering | 0.3 | | (4) |
| 7 | $E_{cont}$ | Effective activation energy for silicate weathering | 20 | kJ mol$^{-1}$ | (5) |
| 7 | $T_s^{PD}$ | Present-day mean surface temperature | 288 | K | |
| 7 | $pCO_2^{PD}$ | Preindustrial atmospheric $pCO_2$ | 280 | ppmv | |
| 7,8 | $R$ | Universal gas constant | 8.31 | J mol$^{-1}$ K$^{-1}$ | |
| 7 | $F_w^{PD}$ | Reference value of continental silicate weathering | 8.05 | Tmol yr$^{-1}$ | (1) |
| 8 | $F_{sw}^{PD}$ | Reference value of seafloor weathering | 0.45 | Tmol yr$^{-1}$ | (1) |
| 8 | $[H^+]_{pore}^{PD}$ | Reference value of hydrogen ion concentration in basalt pore water (Eq. 9) | $1.12 \times 10^{-9}$ | mol L$^{-1}$ | |
| 8 | $T_{pore}^{PD}$ | Reference value of in pore space temperature (Eq. 11) | 283 | K | |
| 8 | $\gamma$ | Exponent for basalt dissolution | 0.25 | | (6) |
| 8 | $E_{diss}$ | Activation energy for basalt weathering | 80 | kJ mol$^{-1}$ | (6) |

References. (1) Catling & Kasting (2017); (2)Cawood et al. (2022); (3) Berner (1994); (4) Walker et al. (1981); (5) Krissansen-Totton & Catling (2017); (6) Krissansen-Totton et al. (2018a).

## 2.5. Experimental setting

We perform a series of numerical experiments under systematically varied planetary and stellar conditions to examine how atmospheric composition and climate respond to key controlling parameters. Specifically, we investigate the response to four parameters: (1) the effective stellar flux, $S_{eff}$, (2) the seafloor spreading rate, $r_{spr}$, (3) stellar spectral type (F-, G-, K-type), and (4) the redox state of volcanic gases, $f_{O2}$. The experimental settings adopted in this study are summarized in Table 3.

We first conduct sensitivity experiments with respect to $S_{eff}$ for the young Sun setting (*YoungSun_$S_{eff}$*). $S_{eff}$ controls the photon flux to the atmosphere and exerts a fundamental control on climate, leading to the response of the global carbon cycle. A change in the surface temperature influences atmospheric chemistry by affecting the partial pressure of water vapor and, thus, number density of OH and H radicals.

We next examine the combined effects of stellar insolation and volcanic outgassing intensity by varying $r_{spr}$ together with $S_{eff}$ (*YoungSun_$S_{eff}$&$r_{spr}$*). Because $r_{spr}$ linearly scales the global volcanic outgassing fluxes, this experiment assesses the influence of volcanic activity on atmospheric composition and climate.

To assess the role of the redox state of volcanic gases, we perform additional experiments in which the oxygen fugacity ($f_{O2}$) of volcanic gases is set to $\log_{10} f_{O2}$ = FMQ − 1, one log unit lower than in the standard experiments (*YoungSun_FMQ-1*; Appendix B1). Variations in $f_{O2}$ affect the relative abundances of reduced and oxidized volcanic gases and thus influence atmospheric redox chemistry (e.g., Holland 2002; Kadoya et al. 2020; Wogan et al. 2020).

Finally, we investigate the sensitivity of our results to stellar spectrum type. For the G-type star, we adopt both the present-day (*G-type*) and 4 Ga (*YoungSun*) solar spectrum (G2V; Thuiller et al. 2004; Claire et al. 2012). We also conduct experiments for early Earth–like planets orbiting F-type (σ Bootis, F2V) (*F-type*) and K-type (ε Eridani, K2V) stars (*K-type*), under otherwise identical conditions (Segura et al. 2003; Arney et al. 2017; Watanabe & Ozaki 2024).

Table 3. List of numerical experiments conducted in this study.

| Experiment name | $S_{\text{eff}}$ | $r_{\text{spr}}$ | $f_{O_2}$ | Stellar spectral type | Figure |
|---|---|---|---|---|---|
| *YoungSun_$S_{\text{eff}}$* | 0.2−1.4 | 10 | FMQ | G2V (Sun, 4 Ga) | 2 |
| *YoungSun_$S_{\text{eff}}$&$r_{\text{spr}}$* | 0.2−1.4 | $10^{-1}$−$10^{1.5}$ | FMQ | G2V (Sun, 4 Ga) | 3 |
| *YoungSun_FMQ−1* | 0.2−1.4 | $10^{-1}$−$10^{1.5}$ | FMQ−1 | G2V (Sun, 4 Ga) | B1 |
| *G-type* | 0.2−1.4 | $10^{-1}$−$10^{1.5}$ | FMQ | G2V (Sun, 0 Ga) | 4, 5, 6 |
| *F-type* | 0.2−1.4 | $10^{-1}$−$10^{1.5}$ | FMQ | F2V (σ Bootis) | 4, 5, 6 |
| *K-type* | 0.2−1.4 | $10^{-1}$−$10^{1.5}$ | FMQ | K2V (ε Eridani) | 4, 5, 6 |

## 3. Results

### 3.1. Response to effective stellar flux

We first examine how climate and atmospheric composition respond to changes in the effective stellar flux, $S_{\text{eff}}$, under a young Sun setting (*YoungSun_$S_{\text{eff}}$*). The coupled climate-carbon cycle model predicts warm climates for $0.28 \leq S_{\text{eff}} \leq 1.10$ (Figure 2a), with surface temperature decreasing monotonically as $S_{\text{eff}}$ decreases. In steady states, volcanic $CO_2$ outgassing must be balanced by silicate weathering (Section 2.4.2). Thus, warmer climates at higher $S_{\text{eff}}$ promote silicate weathering and lower atmospheric $pCO_2$, whereas cooler climates at lower $S_{\text{eff}}$ suppress silicate weathering and require higher atmospheric $pCO_2$ to restore surface temperatures at which silicate weathering can match outgassing. Consequently, atmospheric $pCO_2$ increases systematically with orbital semi-major axis, consistent with previous studies (e.g., Kadoya & Tajika 2014, 2019; Lehmer et al. 2020; Affholder et al. 2025). At $S_{\text{eff}} < 0.70$ atmospheric $pCO_2$ exceeds ~0.3 bar, potentially leading to $CO_2$ condensation and cloud formation (e.g., Caldeira & Kasting 1992; Kadoya & Tajika 2014). Given the complex radiative properties of $CO_2$ clouds (e.g., Mischna et al. 2000; Kitzmann 2016), climate predictions for planets located near the outer edge of the HZ are subject to considerable uncertainty.

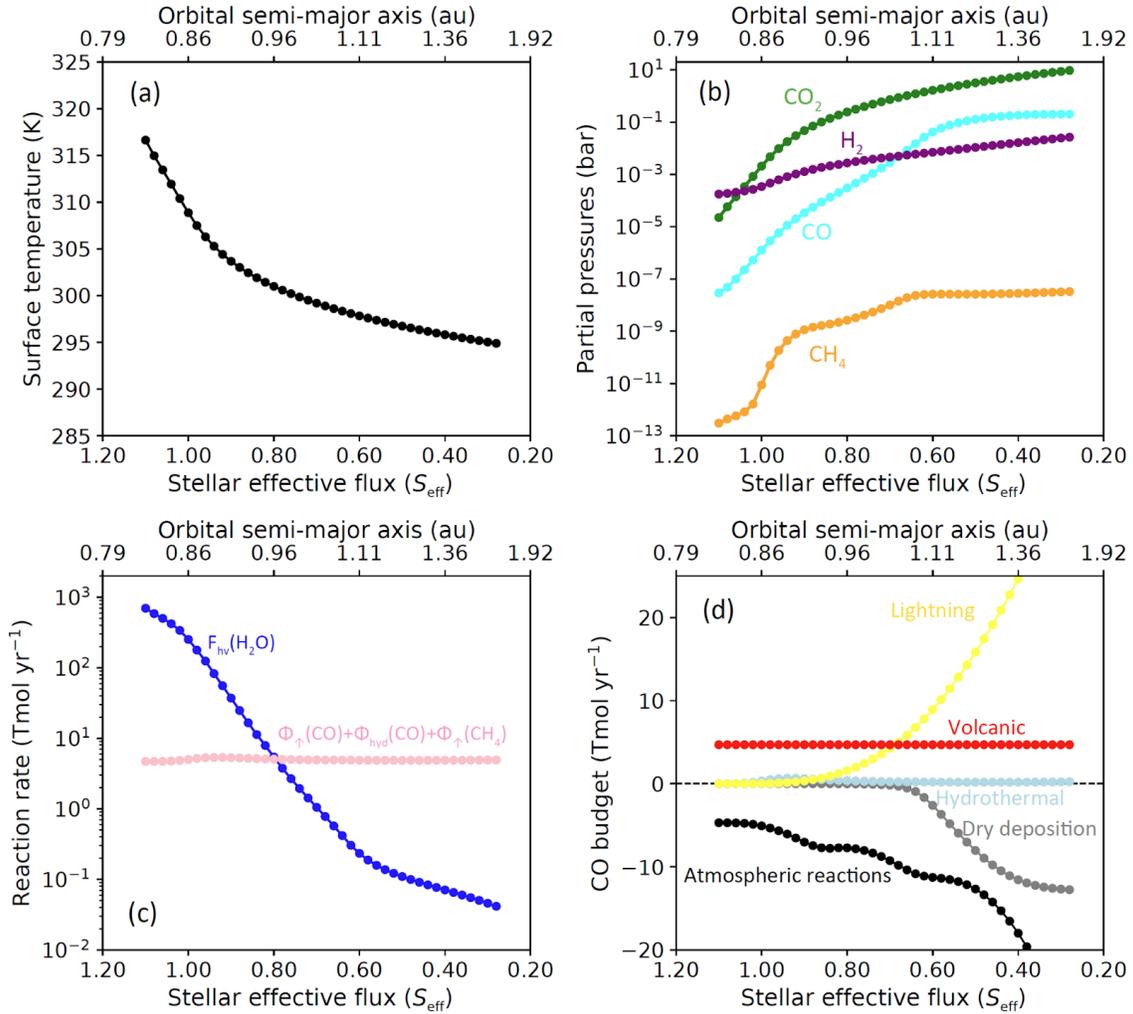

Figure 2. Steady-state response to changes in the effective stellar flux, $S_{\rm eff}$, for the young Sun setting (*YoungSun_$S_{eff}$*). (a) Global mean surface temperature, $T_s$. (b) Atmospheric partial pressures of $CO_2$ (green), CO (cyan), $CH_4$ (orange), and $H_2$ (purple). (c) Photolysis rate of $H_2O$ (blue) and total input rate of reducing carbon species (CO and $CH_4$, pink). (d) CO budget in the atmosphere, where positive and negative values represent source and sink fluxes, respectively. The corresponding orbital semi-major axis is indicated on the top axis. For $S_{\rm eff} < 0.28$ and $S_{\rm eff} > 1.1$, atmospheric $pCO_2$ levels are either too high (>10 bar) or too low (<$10^{-5}$ bar) for the polynomial parameterizations of $J_{\rm OLR}$ and $a$ (Williams & Kasting 1997).

The photochemical model shows that elevated $pCO_2$ enhances CO production via $CO_2$ photolysis ($CO_2 + h\nu$ ($\lambda < 204$ nm) $\rightarrow CO + O$). As a result, $pCO$ increases substantially as $S_{eff}$ decreases (Figure 2b): $pCO$ is as low as $\sim 10^{-8}$ bar near the inner edge of the HZ and increases outward, saturating at $\sim 0.2$ bar around 1.1 au ($S_{eff}= \sim 0.6$). In contrast, $CH_4$ increases with orbital semi-major axis but remains below $\sim 10^{-7}$ bar. The atmospheric $H_2$ abundance shows only a weak dependence on the semi-major axis and remains between $10^{-3}$–$10^{-2}$ bar across the considered range.

The marked increase in $pCO$ at low $S_{eff}$ results from the onset of CO runaway. CO removal is controlled primarily by the availability of OH radicals produced by $H_2O$ photolysis ($H_2O + h\nu$ ($\lambda < 240$ nm) $\rightarrow H + OH$). As $S_{eff}$ decreases, increasing $pCO_2$ enhances UV shielding, reducing the vertically integrated $H_2O$ photolysis rate, $F_{h\nu}(H_2O)$ (Figure 2c). When $F_{h\nu}(H_2O)$ becomes comparable to the total flux of reducing carbon species (CO and $CH_4$), OH production becomes insufficient to oxidize CO ($CO + OH \rightarrow CO_2 + H$), allowing CO to accumulate (Kasting 1990; Watanabe & Ozaki 2024). In the present results, this transition occurs at $S_{eff} = \sim 0.7$.

The atmospheric CO budget further clarifies this transition. At $S_{eff} > 0.66$, the external inputs of CO are balanced mainly by atmospheric reactions, with negligible dry deposition (Figure 2d). As $S_{eff}$ decreases below 0.66, dry deposition becomes an increasingly important sink, indicating a transition toward CO-runaway conditions. This budget-based threshold ($\sim 0.66$) is slightly lower than the threshold inferred from the balance between $H_2O$ photolysis and reduced carbon inputs (Figure 2c), but both consistently identify the onset of CO runaway.

### 3.2. Combined effects of stellar flux and volcanic activity

We next explore the combined effects of $S_{eff}$ and seafloor spreading rate ($r_{spr}$) on planetary climate and atmospheric chemistry (*YoungSun_$S_{eff}$&$r_{spr}$*). The modeled Earth-like planet orbiting a G-type star maintains temperate surface conditions over a broad region of $S_{eff}$-$r_{spr}$ phase space (Figure 3a). Within this temperate regime, surface temperature increases with both $S_{eff}$ and $r_{spr}$.

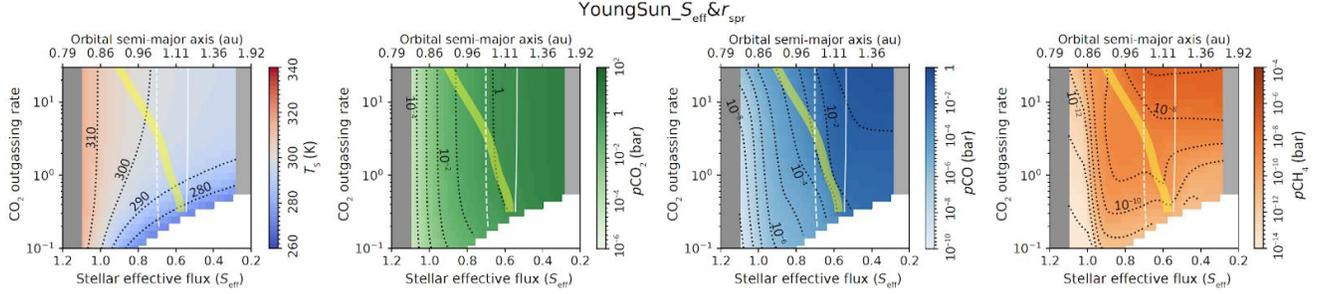

Figure 3. Steady-state response in a phase space of effective stellar flux, $S_{\text{eff}}$, and the seafloor spreading rate, $r_{\text{spr}}$ (*YoungSun_$S_{eff}$&$r_{spr}$*). The y-axis is shown as the corresponding $CO_2$ outgassing flux normalized to the present-day Earth value, rather than the seafloor spreading rate $r_{\text{spr}}$ itself. (a) Global mean surface temperature, $T_s$, (b) $pCO_2$, (c) $pCO$, and (d) $pCH_4$. The yellow line represents the conditions for $F_{\text{hv}}(H_2O) = \Phi_\uparrow(CO) + \Phi_{\text{hyd}}(CO) + \Phi_\uparrow(CH_4)$, corresponding to the onset of CO runaway. The gray-shaded areas represent regions where the climate-carbon cycle model is not applicable ($pCO_2 < 10^{-5}$ bar or $> 10$ bar), and the white area denotes the snowball state. $CO_2$ clouds are expected to form when atmospheric $pCO_2$ exceeds the $CO_2$ saturation vapor pressure given by Caldeira & Kasting (1992, Equation 6). The dashed (poles) and solid (equator) white lines mark these cloud-formation thresholds; $CO_2$ clouds appear at lower $S_{\text{eff}}$ than these lines. Such cases fall outside the strict applicability of the climate-carbon cycle model.

The present results also demonstrate that Earth-like planets may be in a snowball state, particularly under conditions of low $r_{\text{spr}}$ and/or $S_{\text{eff}}$. Within the snowball regime, planets receiving relatively high $S_{\text{eff}}$ (>0.4) are expected to exhibit repeated global glaciation and deglaciation (e.g., Tajika 2007; Kadoya & Tajika 2014; Haqq-Misra et al. 2016). In contrast, at lower $S_{\text{eff}}$ (< 0.4), the climate is expected to remain permanently in a snowball state because of the $CO_2$ condensation. Even within the temperate climate regime, sufficiently low $S_{\text{eff}}$ (< 0.7) may lead to the $CO_2$ condensation. Because $CO_2$ cloud formation and its climatic impacts are not explicitly treated in our climate model, climate states in this region should be considered uncertain.

Atmospheric $pCO_2$ and $pCO$ increase with decreasing $S_{\text{eff}}$ and increasing $r_{\text{spr}}$. CO runaway occurs under both low $S_{\text{eff}}$ and high $r_{\text{spr}}$, consistent with the reduced OH availability under these conditions. Atmospheric $pCH_4$ also increases with decreasing $S_{\text{eff}}$ and increasing $r_{\text{spr}}$. The increase is pronounced at large $S_{\text{eff}}$ (>0.9), but $pCH_4$ remains below 1 ppmv across the modeled range. At lower $S_{\text{eff}}$, volcanic outgassing exerts a dominant control on $pCH_4$. Even under more reducing volcanic conditions (*YoungSun_FMQ-1*; Appendix B1), $p$CH$_4$ remains <1 ppmv. This is because the present study assumes a relatively oxidized planetary redox state comparable to that of the present Earth, which is unfavorable for methane production, and because only methane released by volcanic outgassing is considered. Consequently, the predicted CH$_4$ levels are lower than those of the previous studies of lifeless planetary atmospheres, such as Wogan et al. (2020) and Watanabe & Ozaki (2024), which assume more reducing redox conditions. The present results suggest that CO-rich atmospheres may form on Earth-like planets over a broad range of the HZ.

### 3.3. Influence of stellar spectral type

The stellar spectrum exerts fundamental control on both photochemistry and climate. Because the amount and spectral distribution of stellar radiation received by a planet depend on stellar type and orbital semi-major axis, we use the effective stellar flux as the primary parameter for comparing planets across different host stars. In our simulations, the stellar flux required to maintain temperate climate conditions is lowest for planets orbiting K-type stars and highest for those orbiting F-type stars (Figure 4a, e, i). This trend arises because less massive stars emit more near-infrared radiation, which is more efficiently absorbed by ice and planetary atmospheres (e.g., H$_2$O and CO$_2$), thereby enhancing greenhouse warming (e.g., Kopparapu et al. 2013; Haqq-Misra et al. 2016; Kadoya & Tajika 2019).

The spatial patterns of $p$CO$_2$ in $S_{\text{eff}}$-$r_{\text{spr}}$ space are broadly similar across stellar spectral types. In contrast, $p$CO varies strongly with stellar spectral type: planets orbiting K-type stars tend to have higher $p$CO and exhibit CO runaway over a wider region of their HZs (Figures 4c, g, k). This occurs because K-type stars emit weaker UV flux, reducing H$_2$O photolysis (OH production), thereby limiting atmospheric CO removal (e.g., Ranjan et al. 2022; Watanabe & Ozaki 2024). For planets orbiting F-type stars (Figure 4c), CO runaway can also occur, but the predicted runaway region overlaps with the parameter space where CO$_2$ condensation becomes important (white lines). Therefore, the extent of CO runaway atmospheres for F-type stars should be regarded as uncertain.

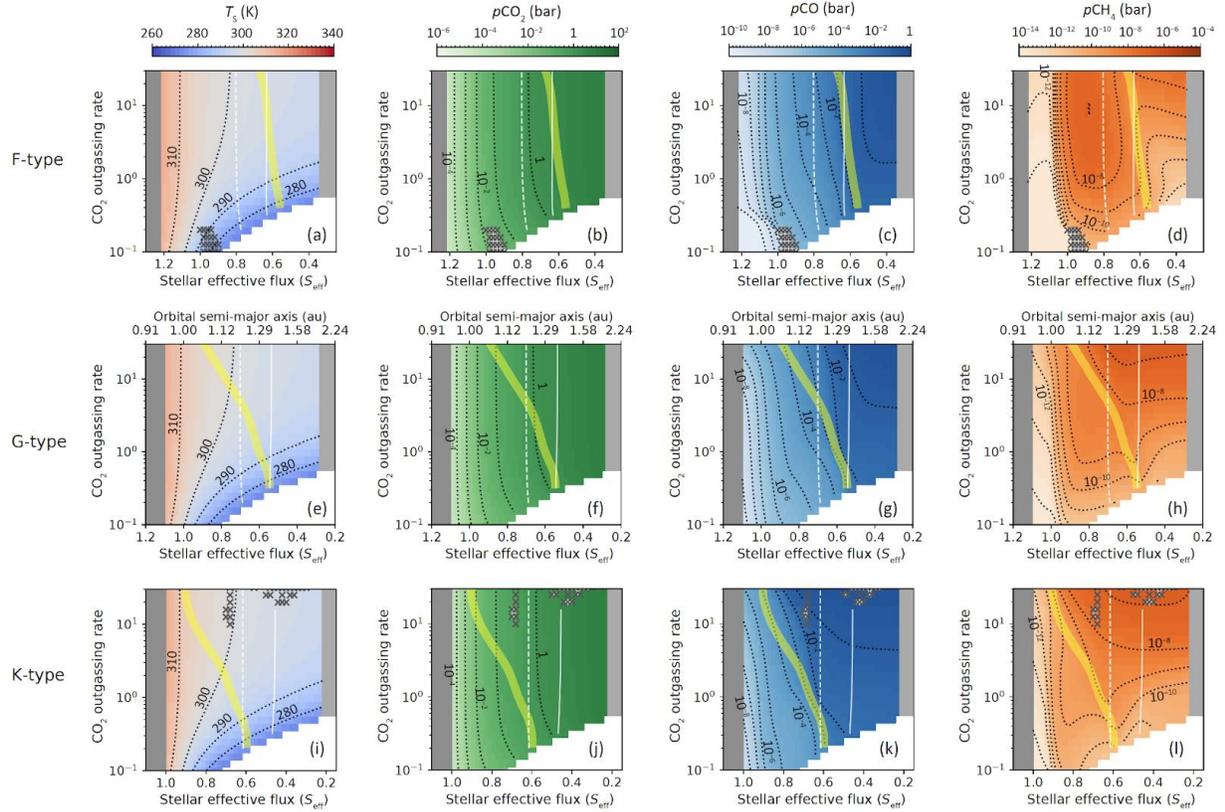

Figure 4. Same as Figure 3, but calculated using the F2V (a-d), present-day Sun (G2V) (e-h), and K2V (i-l) spectra (F-type, G-type, K-type experiments). Note that the ranges of the horizontal axes differ among the stellar-type experiments. X denotes cases for which the photochemical simulation failed to converge.

We also compared simulations using the present-day solar spectrum (Figure 4e-h) with those using the 4.0 Ga solar spectrum (Figure 3). Differences between the two cases are minor because the adopted solar spectral evolution model predicts only modest changes near ~200 nm, the wavelength range most relevant to $H_2O$ photolysis (Figure 6 of Claire et al. 2012). As a result, the stellar-type dependence of $CO_2$, CO, and $CH_4$ is dominated by spectral differences between F-, G-, and K-type stars rather than evolutionary changes in the solar UV output.

Trends in $pCO_2$, $pCO$, and $pCH_4$ within the HZ across stellar spectral types and volcanic outgassing intensities are shown in Figure 5. Overall, $pCO_2$, $pCO$, and $pCH_4$ increase with decreasing $S_{eff}$, over a wide range of outgassing rates and stellar spectral types. In particular, the present results highlight a marked increase in $pCO$ (CO runaway) at low $S_{eff}$ for F-, G-, and K-type stars (Figure 5b, f, j). Methane levels also increase toward lower $S_{eff}$ (Figure 5c, g, k), due to reduced OH radical abundances associated with lower surface temperatures and the resulting decrease in atmospheric $H_2O$ (Grenfell et al. 2007; Akahori et al. 2024), although $pCH_4$ remains a trace gas (<$10^{-6}$ bar). Throughout the parameter space explored here, CO is consistently more abundant than $CH_4$ ($CO/CH_4 > 1$; Figure 5d, h, l). The $CO/CH_4$ ratio reaches a minimum near the inner edge of the CO runaway regime, because $CH_4$ begins to increase at relatively higher $S_{eff}$ than CO. This behavior is especially pronounced for planets orbiting F-type stars (Figure 5d).

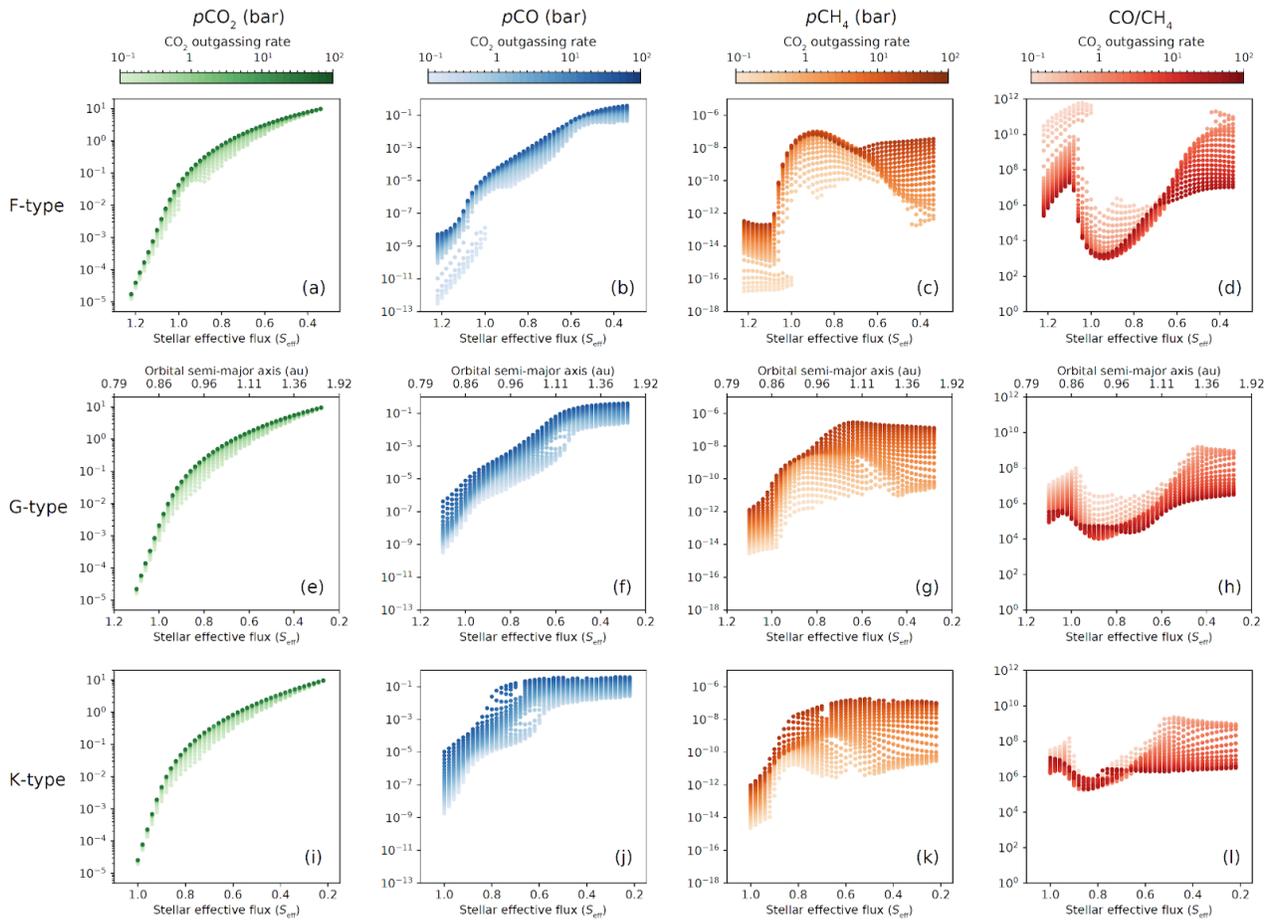

Figure 5. Trends in $pCO_2$, $pCO$, and $pCH_4$, as well as the column-density ratio of $CO/CH_4$ as a function of $S_{eff}$ (or equivalently the orbital semi-major axis) within the HZ. Panels (a–d), (e–h), and (i–l) show results for planets orbiting F-, G-, and K-type stars, respectively (*F-type*, *G-type*, *K-type* experiments). From left to right, columns correspond to $pCO_2$, $pCO$, $pCH_4$, and $CO/CH_4$ ratio. Color intensity indicates the $CO_2$ outgassing rate normalized to the present-day Earth value, with darker shades representing higher outgassing rates. Note that the ranges of the horizontal axes differ among the stellar-type experiments.

## 3.4. Formaldehyde production

In an anoxic prebiotic atmosphere, a variety of organic compounds can form through photochemical processes (e.g., Sagan & Khare 1971; Chyba & Sagan 1992). Among these species, formaldehyde ($H_2CO$) is of particular interest because (i) it is among the major organic photochemical products expected in the atmospheres of the lifeless Earth-like planets considered here, and (ii) polymerization of $H_2CO$ via the formose reaction produces sugars such as ribose that are relevant to the origin of life (e.g., Cleaves 2008; Hashidzume 2025).

Formaldehyde is produced primarily from CO and H radicals via the HCO radical as an intermediate (Pinto et al. 1980):

$$H + CO + M \rightarrow HCO + M, \quad (R1)$$

$$HCO + HCO \rightarrow H_2CO + CO, \quad (R2)$$

where M represents a third molecule. The $H_2CO$ deposition flux exhibits a pronounced peak as a function of $S_{eff}$ (Figure 6a, c, e). More specifically, under G-type star conditions, the peak occurs around $S_{eff}$ of 0.8, with maximum deposition rates on the order of 1 Tmol C yr$^{-1}$. Under CO runaway conditions, however, $H_2CO$ deposition decreases to ~0.1 Tmol C yr$^{-1}$. This behavior reflects the competing roles of H and CO (Watanabe & Ozaki 2024): although $p$CO increases toward the outer edge of the HZ (Figure 3c), the supply of H radicals decreases because $H_2O$ photolysis becomes less efficient due to (i) enhanced $CO_2$ UV shielding, (ii) reduced stellar photon flux, and (iii) lower $H_2O$ vapor abundance in cooler climates. Consequently, $H_2CO$ production peaks at intermediate $S_{eff}$ and at an orbital semi-major axis just interior to the onset of CO runaway. Our results also demonstrate that the stellar spectral type strongly modulates $H_2CO$ production: planets orbiting more massive stars with stronger UV radiation, such as F-type stars, exhibit significantly higher $H_2CO$ deposition rates (Figure 6a). Maximum values reach ~10 Tmol C yr$^{-1}$ for F-type stars, compared to ~0.1 Tmol C yr$^{-1}$ for K-type stars. In all cases, the $S_{eff}$ at which the $H_2CO$ deposition rate peaks lies interior to the CO runaway threshold, similar to the G-type star experiments.

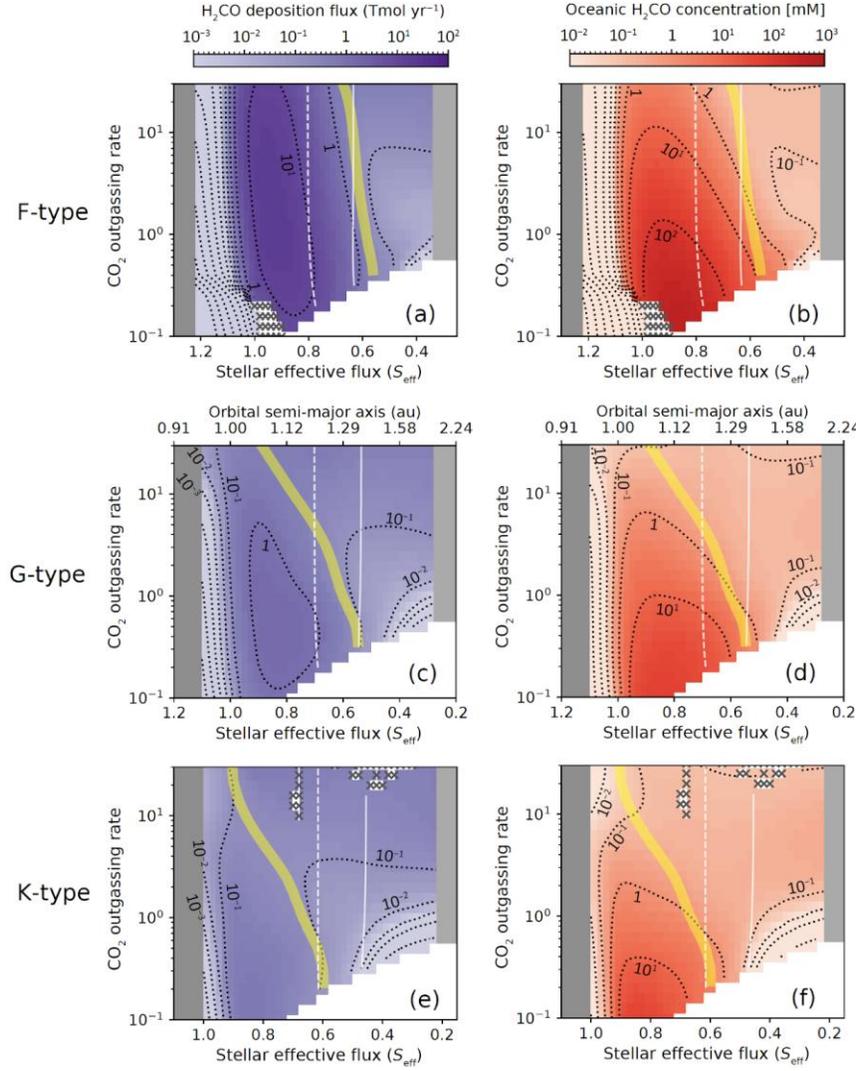

Figure 6. Same as Figures 3 and 4, but showing the steady-state deposition and oceanic accumulation of formaldehyde ($H_2CO$). Panels (a, c, e) show the $H_2CO$ deposition flux to the surface, and panels (b, d, f) show the estimated maximum steady-state concentration of $H_2CO$ in the ocean, for planets orbiting F-, G-, and K-type stars (*F-type*, *G-type*, *K-type* experiments).

Following previous studies (Kharecha et al. 2005; Harman et al. 2013; Watanabe & Ozaki 2024), we estimate the maximum steady-state concentration of $H_2CO$ in the ocean (Figure 6b, d, f). We assume that $H_2CO$ deposited from the atmosphere is removed solely through hydrothermal circulation. Using the present Earth's ocean volume (~$1.4 \times 10^{21}$ L) and the hydrothermal circulation rate at mid-ocean ridges (~$1.4 \times 10^{14}$ L yr$^{-1}$), and scaling the latter with the normalized seafloor spreading rate, $r_{spr}$, we find that the steady-state oceanic concentration of $H_2CO$ generally ranges from ~0.1 to 100 mM, except near the inner edge of HZ. The estimated concentrations follow the same pattern as the deposition flux, peaking at $S_{eff}$ higher than the CO-runaway threshold. For planets orbiting F-type stars, the maximum oceanic $H_2CO$ concentration reaches on the order of 100 mM, whereas for planets orbiting G-type and K-type stars it is typically on the order of 10 mM. As a rough timescale estimate, with a deposition flux of 1 Tmol C yr$^{-1}$, and neglecting losses, it would take around $1.4 \times 10^6$ yr for the ocean to accumulate an $H_2CO$ concentration of 1 mM. This estimate is broadly consistent with the pioneering estimations of Pinto et al. (1980).

## 4. Discussion
### 4.1. Stellar control of CO-runaway atmospheres

Our simulations reveal systematic and physically interpretable trends in the abundances of $CO_2$, CO, and $CH_4$ across the HZ (Figures 4 and 5). As stellar flux decreases, the climate–carbon cycle feedback drives atmospheric $CO_2$ upward, consistent with classical weathering-stabilized climate models (e.g., Kadoya and Tajika 2014; Lehmer et al. 2020). Elevated $CO_2$ then enhances CO production through both photolysis and reduced OH availability, producing a monotonic rise in CO toward the outer edge of the HZ. $CH_4$ shows a weaker increase, limited by its short photochemical lifetime and the modest volcanic fluxes assumed here (cf. Wogan et al. 2020; Watanabe & Ozaki 2024). These coupled trends show that the outer edge of the HZ naturally favors $CO_2$- and CO-rich atmospheres. Importantly, these trends reflect fundamental behavior of the coupled photochemistry–carbon cycle–climate system and are expected to apply broadly to lifeless Earth-like planets within the HZ.

The model also demonstrates that the emergence of CO-runaway atmospheres on lifeless Earth-like planets is fundamentally affected by stellar spectral type. Although CO tends to accumulate under conditions of low insolation and elevated volcanic outgassing across all stellar types, the threshold for runaway behavior varies systematically with stellar mass. CO runaway arises when OH radicals—the dominant sink for CO—become severely depleted. Because near-UV photons drive $H_2O$ photolysis, planets orbiting cooler, lower-mass stars experience persistently lower OH production, enabling CO to build up more readily. Consequently, K-type stars exhibit the largest regions of parameter space in which CO runaway occurs, followed by G-type stars, while F-type stars allow runaway only within a narrow and uncertain range that overlaps conditions conducive to $CO_2$ cloud formation.

Taken together, these results show that CO-runaway atmospheres may be common near the outer edges of the HZs of K- and G-type stars, whereas they are less likely for planets orbiting more UV-luminous F-type stars. The stellar spectral type therefore emerges as a first-order determinant of whether a lifeless Earth-like planet tends toward a CO-rich atmospheric state, providing a natural framework for interpreting both future observations and the statistical occurrence of CO-bearing atmospheres among temperate terrestrial exoplanets.

### 4.2. Implications for prebiotic organic synthesis

The systematic trends of atmospheric chemistry across the HZs of Sun-like stars provide a framework for understanding where prebiotic organic synthesis is most efficient. Our results show that the potential for prebiotic organic synthesis on lifeless Earth-like planets depends sensitively on both orbital distance and stellar spectral type. A central finding of this study is that $H_2CO$ production exhibits a distinct maximum at orbital distances just interior to the onset of CO runaway (Figure 6). This peak arises because key precursors of $H_2CO$ (CO and H) respond differently to decreasing stellar flux: CO increases toward the outer edge of the HZ, whereas H production declines as $H_2O$ photolysis becomes less efficient under lower UV fluxes. The interplay between these opposing trends yields a "prebiotic window" in which both CO and H are sufficiently abundant to sustain efficient $H_2CO$ formation, while CO runaway has not yet been triggered.

The magnitude of $H_2CO$ production within this window varies substantially with stellar spectral type. Planets orbiting more UV-luminous stars, such as F-type stars, achieve the highest $H_2CO$ deposition rates—up to ~10 Tmol C yr$^{-1}$ under favorable outgassing conditions—because abundant near-UV photons sustain high H radical concentrations while still allowing sufficient CO accumulation. In contrast, planets orbiting G-type and K-type stars exhibit progressively lower $H_2CO$ fluxes as stellar UV output decreases. These results indicate that the stellar UV environment, rather than CO availability alone, exerts the dominant control on prebiotic organic synthesis in anoxic atmospheres. Across F-, G-, and K-type stars, $H_2CO$ production peaks at insolations slightly higher than the CO-runaway threshold. This behavior defines a prebiotic region of parameter space in which planets remain CO-poor enough to avoid CO runaway but CO- and H-rich enough to sustain substantial photochemical production of prebiotically relevant organic compounds.

It is often noted, based on classical laboratory experiments on the formose reaction, that $H_2CO$ concentrations of ~100 mM represent a lower limit for sustaining efficient sugar production (e.g., Kitadai & Maruyama 2018; Yadav et al. 2020; Hashidzume 2025). In this context, our results imply that, under the most favorable conditions around F-type stars, oceanic $H_2CO$ concentrations could plausibly reach levels compatible with formose chemistry in global oceans. For G- and K-type stars, or under less favorable conditions even around F-type stars, efficient formose chemistry may require additional localized concentration mechanisms, such as freezing-driven concentration, mineral surface processes, or conversion of formaldehyde into less volatile intermediates (e.g., Cleaves 2008). More recent studies, however, suggest that formose-type reactions may proceed at lower $H_2CO$ concentrations and that other species, such as glycolaldehyde ($C_2H_4O_2$), may instead become the limiting factor (e.g., Ono et al. 2024).

In the present study, we treated $H_2CO$ as a representative prebiotic organic compound, while larger organic molecules produced through atmospheric photochemistry are not explicitly included. The importance of $C_2H_4O_2$ in prebiotic chemistry has been increasingly recognized, as it can act not only as an intermediate but also as a catalyst in the formose reaction (e.g., Powner et al. 2009; Sutherland 2016; Ono et al. 2024). Motivated by this potential relevance, Harman et al. (2013) incorporated $C_2H_4O_2$ into a photochemical model to estimate its atmospheric production. In their study, however, the estimated production rate would result in oceanic concentrations far below those required to sustain the formose reaction. This likely reflects the restricted set of formation pathways considered, which focused on oxidation reactions from ethylene and isoprene (e.g., Bacher et al. 2001). More recent photochemical experiments demonstrate that larger organic molecules, including $C_2H_4O_2$, can also form from CO–$H_2O$ mixtures under UV irradiation (Zang et al. 2025), indicating additional formation mechanisms not previously incorporated into atmospheric models. The formation of $C_2H_4O_2$ and other organic molecules in planetary atmospheres, therefore, represents a promising direction for future laboratory studies and photochemical modeling efforts.

### 4.3. Observational significance of atmospheric CO for exoplanets

With the advent of JWST and the prospect of future missions such as the HWO and the Large Interferometer For Exoplanets (LIFE), atmospheric biosignatures have become a central target in the search for life beyond our planet (National Academies of Sciences 2021; Quanz et al. 2022).

CO is among the most informative gases accessible to current and upcoming facilities because it can accumulate to high abundances under abiotic conditions while being removed by biological activity.

Previous studies have therefore proposed CO could serve as a potential "anti-biosignature" (e.g., Catling et al. 2018). Our results place this idea in a broader planetary context by showing that CO-rich and CO-runaway atmospheres arise naturally on lifeless Earth-like planets, particularly at large orbital semi-major axes and around lower-mass stars such as K-type stars. This stellar-type dependence reflects the reduced near-UV flux of cooler stars, which suppresses OH production and limits atmospheric CO removal.

$CH_4$ has been proposed as a key biosignature gas because most $CH_4$ on modern Earth is biological in origin (e.g., Sagan et al. 1993; Schwieterman et al. 2018; Thompson et al. 2022; Akahori et al. 2024; Seager et al. 2025). However, abiotic processes such as serpentinization, volcanic outgassing, and impacts could also generate significant $CH_4$ fluxes (e.g., Etiope & Lollar 2013; Wogan et al. 2020), making $CH_4$ alone an ambiguous indicator of life. To help resolve this ambiguity, the $CO/CH_4$ ratio has been proposed as a diagnostic tool, since CO is typically co-produced abiotically but efficiently consumed by biology (e.g., Krissansen-Totton et al. 2018b; Thompson et al. 2022). Our simulations show that, under abiotic conditions, most cases yield $CO/CH_4 \gg 1$ across the explored parameter space (Figure 5d, h, l). These results reinforce the robustness of CO and $CO/CH_4$ ratio as diagnostics of abiotic atmospheres (e.g., Schwieterman et al. 2019; Thompson et al. 2022; Tokadjian et al. 2024; Ozaki et al. in press) and extend their applicability across a wide range of stellar and orbital environments.

From an observational perspective, CO has already been detected in the atmospheres of several exoplanets, including hot Jupiters observed by JWST (e.g., Rustamkulov et al. 2023; Esparza-Borges et al. 2023; Grant et al. 2023). Future missions are expected to extend atmospheric observations to Earth-sized rocky planets within the HZ around Sun-like stars. While detecting trace levels of CO may remain challenging, the high CO abundances predicted under CO-runaway conditions—reaching percent-level mixing ratios or higher—should be substantially more accessible. Our results, therefore, motivate targeted searches for CO in planets near the outer regions of the HZ, particularly around K-type stars, where abiotic CO accumulation is most likely. Quantifying the detectability of such CO-rich atmospheres with HWO and LIFE represents an important direction for future observational studies. The detectability of atmospheric CO with these facilities has already been explored in several previous studies. For reflected-light observations with HWO-class missions, CO detection in the 1.55 μm band may be challenging because of overlap with $CO_2$ absorption, although high CO abundances and extended wavelength coverage could improve detectability (e.g., National Academies of Sciences 2021; Tokadjian et al. 2024; Krissansen-Totton et al. 2025). In the thermal infrared, simulations for LIFE suggest that detecting trace levels of CO comparable to modern Earth would be difficult (Konrad et al. 2022; Quanz et al. 2022), but substantially higher CO abundances have not yet been systematically evaluated. In this context, the CO-rich and CO-runaway atmospheres predicted in our simulations represent particularly favorable targets, motivating detectability studies for such extreme but physically plausible abiotic atmospheres.

### 4.4. Implications for early Earth and Mars

Our results indicate that the early Earth's atmosphere was likely close to the threshold for CO-runaway conditions. For early Earth-like conditions (an orbital semi-major axis of 1 au and $S_{eff} \approx$ 0.74 at 4.0 Ga), our model predicts that the threshold $CO_2$ outgassing flux is approximately six times that of present-day Earth (Figure 4c). The outgassing flux is generally thought to have been higher on early Earth because of elevated mantle temperatures (e.g., Sleep & Zahnle 2001; Catling & Kasting 2017). Although the actual outgassing flux remains highly uncertain, even mean outgassing fluxes below this threshold could have been punctuated by transient supply of reducing gases—for instance, during episodes of intense volcanism or following large meteorite impacts—potentially triggering temporary CO-runaway atmospheres (Kasting 1990, 2014; Zahnle et al. 2020; Wogan et al. 2023).

In contrast to early Earth, early Mars likely resided more firmly within the CO-runaway regime in our model framework. At the Martian orbital semi-major axis (1.52 au; $S_{eff} \approx 0.32$ at 4.0 Ga), our simulations indicate that any warm climate state would coincide with conditions favorable for CO runaway (Figure 4c). This reflects the combined effects of low stellar insolation, elevated $pCO_2$ required for climate warming, and limited OH production under weak UV flux. Although the volcanic outgassing flux and surface boundary conditions of early Mars remain highly uncertain, previous studies suggest that maintaining warm conditions would have required high $pCO_2$ and additional greenhouse forcing, conditions that are broadly consistent with a CO-rich atmospheric state (e.g., Batalha et al. 2015; Sholes et al. 2017; Koyama et al. 2024; Adams et al.

2025). These results imply that, if early Mars experienced episodically warm climates, CO-runaway or CO-rich atmospheres may have been a natural consequence under anoxic conditions.

**4.5. Limitations and future direction**

The present study adopts an Earth-centric framework, assuming Earth-like planets with active plate tectonics and a long-term carbon cycle regulated by silicate weathering. The prevalence of such planets in the universe remains uncertain. Nevertheless, within this framework, our model predicts robust and systematic trends in atmospheric composition, including increases in atmospheric $CO_2$, CO, and $CH_4$ with orbital semi-major axes under temperate conditions. These trends provide testable hypotheses that can be evaluated with future observations and comparative studies of terrestrial exoplanet atmospheres.

Our climate calculations neglect the radiative impacts of minor atmospheric species such as $CH_4$, because under conditions explored here planetary climates are dominated by $CO_2$ and $H_2O$ greenhouse forcing. However, under more reducing outgassing scenarios, such as those considered by Wogan et al. (2020) and Watanabe and Ozaki (2024), atmospheric $CH_4$ levels could exceed 100 ppmv and potentially exert non-negligible climatic effects. Investigating such extreme regimes will require fully coupled photochemistry-climate models.

Uncertainties in surface-atmosphere exchange processes also affect our results. In particular, we adopt a constant CO deposition velocity of $1.0 \times 10^{-8}$ cm s$^{-1}$ at a seawater pH of 8, following Kharecha et al. (2005). This rate likely depends on oceanic pH and surface temperature (e.g., Van Trump & Miller 1973), yet remains poorly constrained, particularly under near-neutral pH conditions relevant to early Earth-like oceans. If oceans were more acidic—due to elevated atmospheric $pCO_2$, as predicted for the early Earth (e.g., Tajika & Matsui 1992; Halevy & Bachan 2017; Krissansen-Totton et al. 2018a; Guo & Korenaga 2025)—CO deposition would be less efficient, leading to higher atmospheric CO abundances. Experimental studies quantifying the pH dependence of CO hydration under varying environmental conditions would therefore be valuable. In addition, reductants in seawater, such as dissolved $Fe^{2+}$, could further influence atmospheric redox chemistry. Enhanced reductive capacity in the ocean would decrease the abundance of oxidants such as OH, thereby favoring higher atmospheric $p$CO. Future modeling efforts that explicitly consider redox coupling between the atmosphere and ocean (e.g., Domagal-Goldman et al. 2014) will be crucial for assessing these effects.

The treatment of organic chemistry is also simplified in this study. We focus on $H_2CO$ as a representative prebiotic organic compound and do not explicitly include the formation and destruction of more complex organics containing two or more carbon atoms. Given the possible importance of $C_2H_4O_2$ in prebiotic chemistry (e.g., Powner et al. 2009; Sutherland 2016; Ono et al. 2024), a quantitative and mechanistic understanding of the formation chemistry of $C_2H_4O_2$ in anoxic atmospheres, as well as its degradation and consumption in oceans, represents an important direction for future laboratory, theoretical, and modeling studies.

Finally, while this study focuses exclusively on abiotic planetary environments, incorporating biological processes will be essential for assessing the diagnostic power of atmospheric CO and related species. Previous studies have shown that biological activity can strongly alter atmospheric redox balance and carbon cycling (e.g., Kharecha et al. 2005; Gebauer et al. 2017; Ozaki et al. 2018; Sauterey et al. 2020; Watanabe et al. 2023; Akahori et al. 2024; Eager-Nash et al. 2024). Extending the integrated modeling framework developed here to include biological processes across a broad range of stellar and planetary conditions will be a critical step toward interpreting future observations of potentially inhabited worlds.

## 5. Conclusions

In this study, we developed an integrated atmosphere-climate-carbon cycle framework to investigate the atmospheric compositions of lifeless, Earth-like planets across the HZs of F-, G-, and K-type stars. By explicitly coupling volcanic outgassing, silicate weathering, climate, and photochemistry, we quantified how atmospheric $CO_2$, CO, and $CH_4$ vary with stellar spectral type and orbital distance within the HZs. Our simulations reveal a fundamental trend in carbon chemistry on habitable (but uninhabited) terrestrial planets. Around cooler, lower-mass stars, weaker near-UV radiation suppresses OH production and favors the accumulation of CO, allowing CO runaway atmospheres to develop over a broad region of the HZ. In stark contrast, the atmospheric production and surface deposition of formaldehyde, a key precursor for prebiotic organic chemistry, are most efficient around more UV-luminous stars and peak at orbital distances just interior to the CO-runaway threshold. Taken together, stellar spectral type exerts a first-order control on whether lifeless Earth-like planets develop CO-rich atmospheres or prebiotically active environments.

More broadly, this work establishes a quantitative link between observable system properties—such as stellar spectral type and orbital distance—and the coupled atmospheric carbon chemistry, climate, and carbon cycle of terrestrial planets. By connecting $CO_2$–$CO$–$CH_4$ abundances, CO-runaway behavior, and prebiotic organic synthesis within a unified physical framework, our results provide a basis for interpreting future spectroscopic observations of the atmospheres of Earth-sized exoplanets and for assessing where temperate terrestrial planets are most likely to host chemically favorable conditions for the origin of life.

## Acknowledgments

This research is funded by JSPS KAKENHI Grant Number 22H05149, 22H05150, and the Astrobiology Center Program of National Institutes of Natural Sciences (NINS) (grant number AB0604).

# Appendix A. Parameterization of the surface albedo

The surface albedo, $a_s$, is parameterized as an area-weighted average of the albedos of the land, ocean, ice, and clouds (e.g., Rosing et al. 2010; Kadoya & Tajika 2019):

$$a_s = (1 - f_{cloud})(f_{land}\, a_{land} + f_{ocean}\, a_{ocean} + f_{ice}\, a_{ice}) + f_{cloud}\, a_{cloud}, \quad (A.1)$$

where $a_{land}$, $a_{ocean}$, $a_{ice}$, and $a_{cloud}$ represent the albedos of land, ocean, ice, and clouds, respectively. The corresponding fractional areas satisfy $f_{land} + f_{ocean} + f_{ice} = 1$. Clouds are treated as an areal mask that overrides the underlying surface type. The global cloud fraction is fixed at 0.5, independent of $T_s$, the stellar spectral type, or land–ocean distribution (e.g., Kadoya & Tajika 2019). The land and ocean albedos are tuned so that the model reproduces the present-day $T_s$ of 288 K. Under the present-day Earth conditions, ice covers approximately 8% of the surface (~5% ocean ice and ~3% land ice; Hartmann 2016; Parkinson 2004), yielding a present-day surface albedo of $a_s^{PD} = 0.297$.

To capture the first-order ice-albedo feedback, we distinguish three climate regimes based on the global mean surface temperature $T_s$: (i) ice-free state ($T_s \geq T_0$), (ii) a partially ice-covered state ($T_i < T_s \leq T_0$), and (iii) a globally glaciated (snowball) state ($T_s \leq T_i$). We adopt $T_0 = 300$ K as the lower bound of the ice-free regime, based on estimates of the warm interval preceding Antarctic glaciation (Judd et al. 2024). We also adopt $T_i = 273$ K as the snowball threshold (e.g., Pierrehumbert et al. 2011; Ozaki & Reinhard 2021).

For ice-free conditions, the ice fraction is zero, and the surface fractions are given by $f_{land} = f_{cont}$, $f_{ocean} = 1 - f_a$, yielding

$$a_s = (1 - f_{cloud})[f_{cont}\, a_{land} + (1 - f_{cont})\, a_{ocean}] + f_{cloud}\, a_{cloud}, \quad (A.2)$$

which we denote as $a_s^{warm}$.

In the globally glaciated (snowball) conditions, all clear-sky surfaces are ice-covered:

$$a_s = (1 - f_{cloud})\, a_{ice} + f_{cloud}\, a_{cloud}. \quad (A.3)$$

For partially ice-covered states, the surface albedo is linearly interpolated between $T_0$ and 288 K:

$$a_s(T_s) = (a_s^{warm} - a_s^{288K}) \cdot (T_s - T_0) / (T_0 - 288K) + a_s^{warm}, \quad (A.4)$$

where $a_s^{288K}$ represents the surface albedo at 288 K. At 288 K, we assume that both ocean and land ice are present, as on the present Earth, and that the ice-covered areas scale with the clear-sky ocean and land fractions. This parameterization implies that planets with smaller land fractions develop reduced ice-cover at fixed surface temperature, consistent with general circulation model results showing reduced polar ice on ocean-dominated planets (Jenkins et al. 1993). Under these assumptions:

$$f_{\text{ice(land)}}^{288K} = (f_{\text{cont}}/f_{\text{cont}}^{PD}) f_{\text{ice(land)}}^{PD}, \quad (A.5)$$

$$f_{\text{ice(ocean)}}^{288K} = [(1 - f_{\text{cont}}) / (1 - f_{\text{cont}}^{PD})] f_{\text{ice(ocean)}}^{PD}, \quad (A.6)$$

and the total clear-sky ice fraction is:

$$f_{\text{ice}}^{288K} = f_{\text{ice(ocean)}}^{288K} + f_{\text{ice(land)}}^{288K}. \quad (A.7)$$

The corresponding clear-sky ocean and land fractions are:

$$f_{\text{land}}^{288K} = f_{\text{cont}} - f_{\text{ice(land)}}^{288K}, \quad (A.8)$$

$$f_{\text{ocean}}^{288K} = 1 - f_{\text{cont}} - f_{\text{ice(ocean)}}^{288K}. \quad (A.9)$$

The ice albedo depends on the stellar spectral energy distribution. Following Haqq-Misra et al. (2016) and Kadoya & Tajika (2019), we calculate $a_{\text{ice}}$ as a weighted average of visible and near-infrared reflectivities using stellar spectra for F-, G-, and K-type stars. The resulting values are summarized in Table A1. Under early Earth condition (land fraction $f_a = 0.075$), the surface albedo $a_s$ is 0.282 (assuming $T_s = 288$ K). Using the polynomial expressions of Haqq-Misra et al. (2016), the corresponding planetary albedo $a$ is 0.304 for present-day Earth and 0.295 for early Earth.

Table A1. Parameters used for the surface albedo and fractional surface areas.

| Parameter | Value | Ref. |
|---|---|---|
| $a_{\text{ocean}}$ | 0.0745 | (1) |
| $a_{\text{land}}$ | 0.2 | (2) |
| $a_{\text{ice}}$ (F-type host star) | 0.701 | (3) |
| $a_{\text{ice}}$ (G-type host star) | 0.656 | (3) |
| $a_{\text{ice}}$ (K-type host star) | 0.591 | (3) |
| $a_{\text{cloud}}$ | 0.44 | (4) |
| $T_0$ | 300 K | This study |
| $T_i$ | 273 K | This study |
| $f_{\text{cloud}}$ | 0.5 | This study |
| $f_{\text{ocean}}^{PD}$ | 0.65 | This study |
| $f_{\text{land}}^{PD}$ | 0.27 | This study |
| $f_{\text{ice}}^{PD}$ | 0.08 | This study |
| $f_{\text{ice(ocean)}}^{PD}$ | 0.05 | (5) |
| $f_{\text{ice(land)}}^{PD}$ | 0.03 | (6) |

References. (1) Jin et al. (2004); (2) Rushby et al. (2019); (3) Kadoya & Tajika (2019); (4) Cess et al. (1976); (5) Hartmann (2016); (6) Parkinson (2004).

## Appendix B. Results for cases with more reducing volcanic outgassing

In the main text (Figures 4 and 6), the thermodynamic outgassing model assumes that volcanic gases are buffered at an oxygen fugacity of $\log_{10} f_{O2}$ = FMQ. To assess the sensitivity of our results to the redox state of volcanic outgassing, we conducted additional experiments for a more reducing case, in which volcanic gases are buffered at $\log_{10} f_{O2}$ = FMQ − 1 (*YoungSun_FMQ-1*). Overall, the qualitative behavior of the climate-atmosphere system remains unchanged compared with the FMQ case: atmospheric $pCO_2$ and $pCO$ systematically increase toward larger orbital semi-major axes (lower insolation), $pCH_4$ and $pH_2$ increase with normalized outgassing flux, and the $H_2CO$ deposition flux exhibits a distinct peak as a function of orbital semi-major axis.

Quantitatively, however, more reducing outgassing leads to systematically higher abundances of reduced gases. For a given orbital semi-major axis and normalized outgassing flux, $pCH_4$ is typically increased by approximately 1–2 orders of magnitude, compared with the FMQ case. The threshold for CO runaway (yellow curve in Figure B1) shifts inward by ≈0.04 au, indicating that CO runaway occurs at slightly higher stellar fluxes under more reducing outgassing conditions. In addition, the region where the $H_2CO$ deposition flux exceeds 1 Tmol C yr$^{-1}$ expands relative to Figure 4, covering a wider range of orbital semi-major axes and outgassing fluxes.

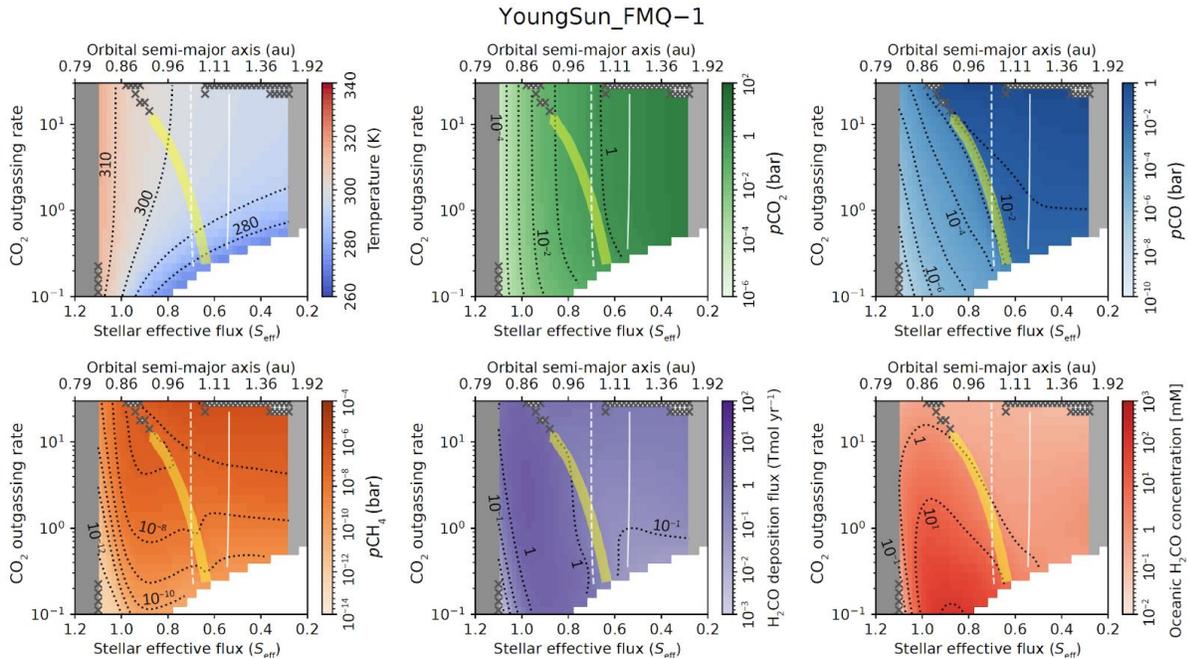

Figure B1. Panels (a–d) and (e–f) are the same as Figures 3 and 6, respectively, but for more reducing volcanic outgassing in the thermodynamic model ($\log_{10} f_{O2}$ = FMQ − 1).